\documentclass[prb,letterpaper,superscriptaddress,aps,floatfix,twocolumn,longbibliography]{revtex4-1}
\usepackage{graphicx}
\usepackage{amsmath}
\usepackage{amsfonts}
\usepackage[small,bf]{subfigure}
\usepackage[colorlinks=true,citecolor=blue]{hyperref}

\usepackage{xcolor}
\usepackage{color}
\usepackage{url}

\begin{document}
\title{Theory of edge states in graphene-like systems}
\author{ J. L. Lado}
\affiliation{Department of Applied Physics, Aalto University, 02150 Espoo, Finland}
\author{J. Fern\'andez-Rossier}
\affiliation{Theory of Quantum Nanostructures Group, International Iberian Nanotechnology Laboratory, 
Av. Mestre Jos\'{e} Veiga, 4715-330 Braga, Portugal\footnote{On permanent leave from Departamento de F\'isica Aplicada, Universidad de Alicante,Spain}}
\date{\today}
\begin{abstract}
Systems that can be described with the same mathematical models that account for the properties of electrons in graphene are known as graphene-like systems. These include magnons, photons, polaritons, acoustic waves, and electrons in honeycomb lattices, either natural or artificial. All of them feature an outstanding property,   the existence of states localized at the edges.  
 We distinguish two classes of edge states depending on whether or not a topological band gap is present in the two-dimensional energy bands. We introduce the theory of edge states in terms of tight-binding models and discuss their properties, with an emphasis on the interplay between their one-way character and the topological nature of the bulk phase.

\end{abstract}
\maketitle
\section*{Keywords}

Bulk-boundary correspondance,
Chern insulator,
Dirac electrons,
Graphene, 
Graphene ribbon,
Haldane model,
honeycomb-lattice, 
Kane-Mele model,
Quantum Spin Hall,
Tight Binding
Topological, 
Two-dimensional crystals,
Zak phase

\section*{Key points}
\begin{itemize}
\item Graphene-like systems can be modeled with a tight-binding model that describes a honeycomb lattice with one state per site, resulting
in peculiar energy dispersions, both  in bulk, with Dirac cones, and at the edges.
\item States localized at the boundaries are known as edge states. 
\item The properties of the edge states are intimately related to  the topological properties of the bulk.  When a topological gap opens in bulk, in-gap edge states exist in all  boundaries. Otherwise,  zero-energy edge states exist only along certain crystallographic directions.
 
\end{itemize}

\section*{Graphene and graphene-like systems, and their edge states}
\label{sec:1}

\subsection*{Main properties of graphene band structure}
Graphene is a two-dimensional material made of carbons forming a honeycomb lattice. Its electronic properties are often described\cite{neto09} with a tight-binding model with only one atomic orbital per carbon and first-neighbour hopping. This  model has  three outstanding  properties:
\begin{enumerate}
\item It  has two energy bands, on the account of the two sites in the unit cell of the honeycomb lattice.  
\item The two energy bands become degenerate at two points in the six corners of the hexagonal Brillouin zone.  
\item In the neighbourhood of these degenerate points, the energy bands form two Dirac cones with electron-hole symmetry.  
\end{enumerate}

In the rest of this chapter, we choose our energy origin, $E=0$, at the degeneracy points and we refer to them as the Dirac points. The lattice shows 
$C_3$ symmetry, that remains invariant under rotations of 120$^\circ$.
Only two groups of 3 equivalent corners  or Dirac points are actually different states of the crystal.  We refer to these groups as {\em valleys}, and we label them as $K$ and $K'$.   

Importantly, there are several physically relevant perturbations to the first-neighbour hopping tight-binding model that lead to the opening of a {\em gap} between the two bands.  In some cases, these perturbations make graphene enter a topological phase, different from conventional insulators, having an important impact on the edge states.  We shall distinguish two broad types of topological states in graphene, Chern insulators\cite{haldane88} and Quantum Spin Hall insulators\cite{kane05}.  The Quantum Hall phase promoted in graphene by the application of an external magnetic field perpendicular to the honeycomb plane\cite{neto09} belongs to the Chern-insulator type. 

\subsection*{Graphene-like systems}

Many of  these properties, including the Dirac cones  in the bulk-dispersion,  the topological-gap openings,  are shared by a class of systems that can be described with the same mathematical models than graphene and its topological variants\cite{polini13,lado15b}. The relevant  degrees of freedom in these graphene-like systems  are not only electrons, but also  magnons,  photons of various wavelengths, polaritons, and sound waves.  Some of these graphene-like systems are artificial arrays of various different types and length-scales: atomically-precise lattices crafted with a scanning tunneling microscope on a surface, arrays of quantum dots in silicon, milimeter size dielectric resonators arranged in a honeycomb lattice, and cm-scale acoustic resonators.  The list of graphene-like systems,  either topological or not,  includes:
\begin{itemize}
\item  Electrons  other 2D materials with honeycomb lattices and sp$^2$, such as silicene and germanene, in  and nanofabricated  honeycomb lattices\cite{singha11}, surfaces with arrays of CO molecules\cite{gomes12}  and molecular crystals
\item  Magnons in ferromagnetic crystals where the magnetic atoms are disposed in a honeycomb lattice \cite{owerre16} such as CrI$_3$\cite{huang17} 
\item  Cold-atoms trapped in tunable optical lattices with honeycomb geometry\cite{tarruell2012}
\item Exciton polaritons in honeycomb arrays of micropillars\cite{jacqmin14}

\item Photons in honeycomb photonic lattices\cite{plotnik2013}

\item Acoustic waves in honeycomb lattices of milimeter scale cylinders\cite{torrent12,khanikaev15}

\end{itemize}

This  list illustrates the diversity of physical systems, length-scales  and degrees of freedom that can be described in terms of the same mathematical models.

\subsection*{Edges and edge states}


Edge states  can occur at the  boundaries of graphene, where carbon atoms have only two carbon first neighbours, instead of three.
They are localized in the direction perpendicular to the boundary and can be extended along the boundary.
The boundaries of   semi-infinite graphene are  one-dimensional (1D) and they can
host  one-dimensional extended edge states.   Very often, the energy of edge states is in the neighbourhood of the Dirac points.   

\begin{figure}
 \centering
    \includegraphics[width=0.5\textwidth]{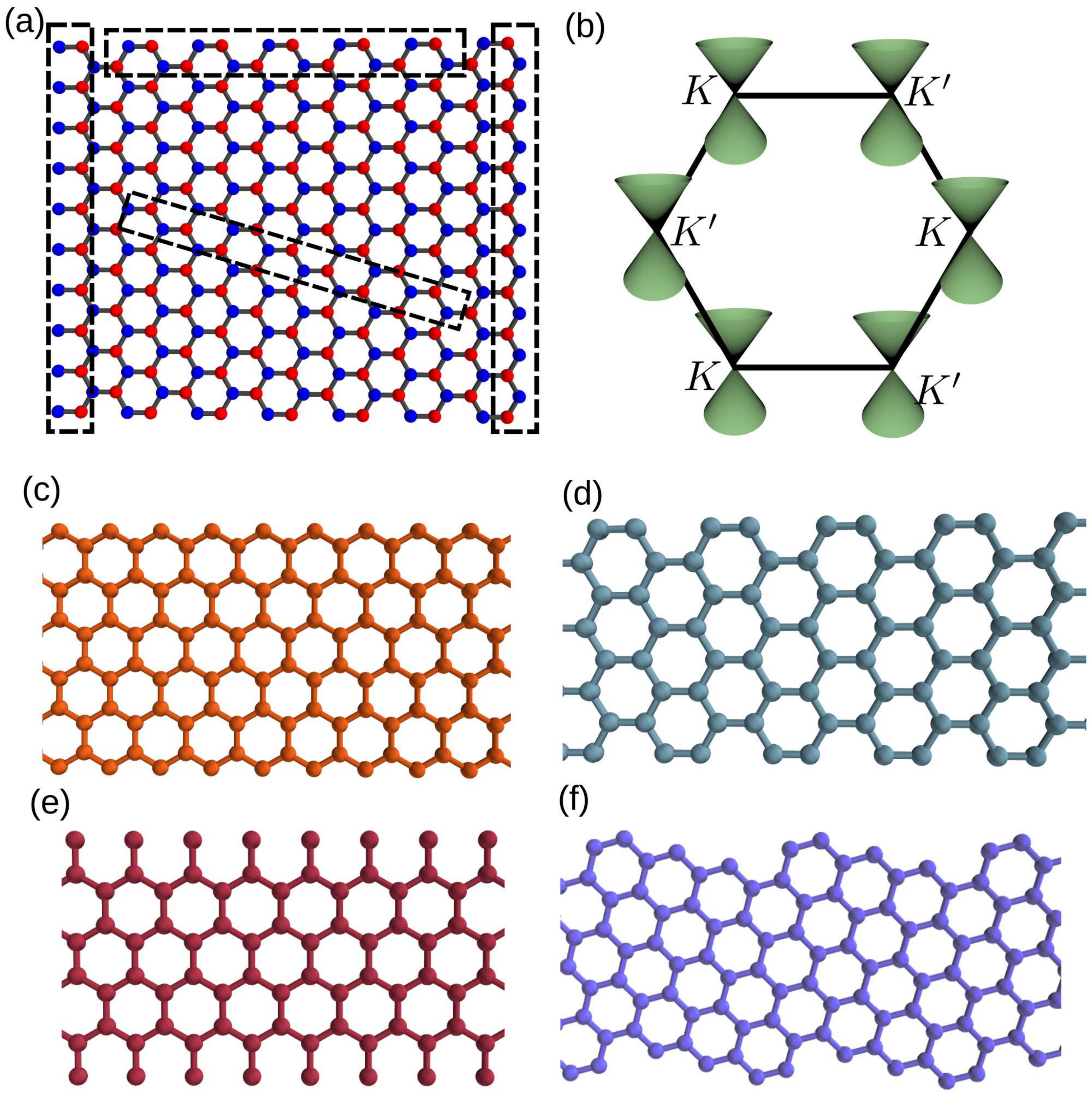}
\caption{(a) Honeycomb lattice, with the two triangular sublattices, $A$ and $B$, are displayed with fake color, red and blue.  (b) Brillouin zone associated to the honeycomb lattice, including the plot of the two  energy bands forming Dirac cones  in the neighborhood of $K$ and $K'$ points (see text).
Four types of edges, zigzag (c), armchair (d), bearded (e) and chiral (f).
}
\label{f1}
\label{fig:structures}
\end{figure}

Edges in graphene-like systems are characterized by their orientation relative to the two high-symmetry directions in the honeycomb lattice, armchair and zigzag (see Fig. 1a,c,d).  The most studied edges are parallel to one of these two directions. However, edges along arbitrary directions, known as chiral edges, can be considered (see Fig.  \ref{f1}f)

We distinguish two types of edge states, depending on whether or not graphene has a topological gap. For gapless graphene,  edge states only occur for edges in a range of orientations. For instance,   zigzag edges host edge states whereas armchair do not. In gapless graphene, edge states have strictly $E=0$ energy forming flat 1D bands\cite{nakada96}.

In the presence of a topological gap in the bands of 2D crystal, all boundaries of graphene do host edge states, and they have dispersive energy bands $E(k)$. For Quantum Spin Hall insulators edge states have spin-momentum locking: at a given edge electrons with opposite spin projection along the off-plane direction  propagate in opposite directions\cite{kane05}.  Thus, back-scattering entails a spin-flip.   In the case of Quantum Hall and Chern insulators, all edge states are chiral:  propagation only occurs in one direction,  regardless of the spin, making back-scattering strictly forbidden\cite{haldane88,lado15b}. 

\section*{Theory of quantum states in  2D honeycomb lattices}
\label{sec:TB}
\subsection{Graphene-like states in two dimensions}

The honeycomb lattice of 2D 
graphene can be treated as  two interpenetrating  triangular lattices,  that we label as $A$ and $B$ in Fig. \ref{fig:structures}a. 
Thus, the honeycomb lattice is a triangular lattice  with two atoms per unit cell that naturally leads, within the simple TB model,  to this  $2\times2$ Bloch Hamiltonian\cite{neto09}:
\begin{eqnarray}
{\cal H}_0(\vec{k})= t\left(
\begin{array}{cc}
0 &  f(\vec{k}) \\
 f^*(\vec{k}) & 0
\end{array}
\right)
\label{H0}
\end{eqnarray}
where  $t$ is the first neighbour hopping and $f(\vec{k})=1+e^{i\vec{k}\cdot\vec{a}_1}+e^{i\vec{k}\cdot\vec{a}_2}$
is the Bloch hopping associated to first neighbour hopping in the honeycomb lattice, and 
\begin{eqnarray}
\vec{a}_1=\frac{a}{2}\left(\sqrt{3},1\right)=a\left(\cos 30^{\circ},\sin 30^{\circ}\right) \nonumber\\
\vec{a}_2=\frac{a}{2}\left(\sqrt{3},-1\right)=a\left(\cos 30^{\circ},-\sin30^{\circ}\right)
\end{eqnarray}
where $a$ is the unit cell spacing, which coincides with the second neighbour distance and it satisfies the relation $a=\sqrt{3}a_{CC}$ with  the first neighbour distance. 
The resulting energy bands, $\epsilon_{\pm}(\vec{k}) = \pm |t| \sqrt{| f(\vec{k}) |^2}$ are shown at low energies in
Fig. \ref{f1}b.  The function $f(\vec{k})$   vanishes at the $K$ and $K'$ points,  so that the two bands meet at those points, with energy $\epsilon=0$.   Due to the bipartite character of the honeycomb lattice, the  energy bands have the so called {\em electron-hole} symmetry, {\em i.e.} for every state with energy $+E$, there is another with energy $-E$.

At this point,  we introduce the concept of sub-lattice as a pseudo spin
degree of freedom, with Pauli matrices $\sigma_a$, $a=x,y,z$,  such that $\sigma_z$ is the sublattice operator.
By so doing,  we can write down the Bloch
Hamiltonian in terms of the Pauli matrices $\sigma_x,\sigma_y$:
\begin{eqnarray}
{\cal H}_0(\vec{k})= 
t \left [ Re(f(\vec{k})) \sigma_x +
Im(f(\vec{k})) \sigma_y \right ]
\end{eqnarray}
The resulting Bloch states can be found to be:
\begin{equation}
\psi_{\vec{k},\nu=\pm 1}= e^{i\vec{k}\cdot\vec{r}} 
\left(\begin{array}{c} 1 \\ \nu e^{i\phi(\vec{k})} \end{array}\right)
\end{equation}
where $\phi(\vec{k})$ is the phase of the complex number $f(\vec{k})$.
Importantly, all finite energy the eigenstates of ${\cal H}_0$, and any other bipartite model,  have an equal weight on both sublattices.

In order to describe the low-energy physics of graphene it is enough to expand the Bloch Hamiltonian of eq. (\ref{H0})
 the neighbourhood of the $K$ and $K'$ points,  denoted by the label $\tau_z=\pm 1$.
 By so doing we obtain:
 \begin{eqnarray}
{\cal H}_{0}(\vec{q})= \hbar v_F \left(q_x \sigma_x  + \tau_z q_y \sigma_y\right) 
\label{kp}
\end{eqnarray}
where $\vec{q}\equiv \vec{k}- \vec{K}_{\tau}$ and $\hbar v_F=\frac{3 }{2}t a_{CC} $, where $a_{CC}$ is the first-neighbour distance 
in the honeycomb lattice.   We thus see that, in the neighbourhood of the $K$ and $K'$ point the graphene tight-binding Hamiltonian of  eq. (\ref{H0}) can be mapped into two copies, one per valley, of a Dirac Hamiltonian.  Many of the low energy properties of graphene-like systems can be understood in terms of the Dirac Hamiltonian.

\subsection*{Band-gap opening}
We now discuss different extra perturbations in the model of eq. (\ref{H0})  that, while preserving the overall structure of the graphene bands, lead to band-gap openings, that may or may not be topological. We start with the simplest case of a non-topological band-gap opening induced by a sublattice symmetry breaking term, as it would happen for instance in 2D boron nitride:
\begin{eqnarray}
{\cal H}_{\rm BN} =
 \left(
\begin{array}{cc}
\frac{\Delta}{2}  &  t\,f(\vec{k}) \\
 t\,f^*(\vec{k}) & -\frac{\Delta}{2}
\end{array}
\right)
\label{BN}
\end{eqnarray}
It is apparent that, at the Dirac points, the off-diagonal terms vanish and the eigenvalues are $E=\pm \frac{\Delta}{2}$, so that a band-gap of dimension $\Delta$ opens-up at the Dirac points. Importantly, $\Delta$ is momentum independent, so that it has the same sign at both valleys.

 We now consider the Haldane model\cite{haldane88} that also leads to a gap opening at the valleys, but with opposite sign. This turns out to
 be essential to make this gap topological. The Haldane model amounts to add to the eq. (\ref{H0}) a term  that accounts for 
local magnetic fields with a vanishing total  flux. 
These magnetic fields give rise to an  imaginary second-neighbour hopping:
\begin{eqnarray}
{\cal H}_{\rm Haldane} =
 \left(
\begin{array}{cc}
t_H g(\vec{k})  &  t\,f(\vec{k}) \\
 t\,f^*(\vec{k}) & -t_H g(\vec{k})
\end{array}
\right)
\end{eqnarray}
where $g(\vec k)= 2	\left(\sin \phi_1 +  \sin\phi_2 +  \sin(\phi_1-\phi_2)\right)$, where $\phi_{1,2}\equiv \vec{k}\cdot\vec{a}_{1,2}$.
In graphene,  the magnetic field required to realize the Haldane term can not be 
directly created externally on account of their very rapid space modulation. In contrast,  in some graphene-like systems, the second-neighbour Haldane hopping can be engineered in the lab. For instance, in the case of magnons, second-neighbour  Dzyaloshinskii-Moriya interactions give rise to Haldane-type terms in the magnon Hamiltonian\cite{owerre16}.  

A $kp$ expansion of the Haldane term around the Dirac point would yield the following extra term in the Dirac Hamiltonian:
\begin{equation}
\delta  {\cal H} = \frac{3}{2} \sqrt{3}t_H \tau_z \sigma_z=\frac{\Delta}{2}\tau_z\sigma_z
\label{HH2}
\end{equation}
so that a gap  of magnitude $\Delta=3\sqrt{3}t_H$ opens-up at the Dirac point.  If we define the {\em sign} of the gap, encoded in $\tau_z$, the gap in the two valleys of the Haldane model has opposite signs.

 This gap turns out to be topological. To see that, we define the Chern number associated to a  band $n$ as
the integral over the Brillouin zone
\begin{equation}
{\cal C}_n =\frac{1}{2\pi} \int_{BZ} \Omega_n d^2k
\label{Chern}
\end{equation}
where  $\Omega_n$  is the Berry curvature
\begin{equation}
\Omega_n =  i
\left (
\partial_{k_x} \langle \Psi_n | \partial_{k_y} | \Psi_n \rangle
-
\partial_{k_y} \langle \Psi_n | \partial_{k_x} | \Psi_n \rangle
\right )
\end{equation}
These expressions, valid for non-degenerate bands, can be easily generalized to the multiband case.

The Chern number can only take integer values
when the integral encompasses the whole Brillouin zone, 
it is  quantized.  Parametric deformations of a Hamiltonian that preserve the
 symmetry and do not close the gap  do not change the Chern number. This robustness is the key property of topological insulators.  Importantly, the value of the Chern number determines
 the Hall conductance via the TKNN formula, 
$\sigma_{xy} = \frac{e^2}{\hbar}\sum_n {\cal C}_n f_n $, where $f_n=0,1$ marks the occupation of the band $n$ at zero temperature.   For trivial two-dimensional insulators, the Chern number vanishes, and so it does the Hall conductance. For a class of  two-dimensional topological insulators, including graphene and other 2D systems in the Quantum Hall regime, the Chern number is finite and the Hall conductance is quantized.    The Haldane model was the first example of a band insulator with non-vanishing Chern number and no Landau Levels.
The existence of  edge states in topological insulators is essential for  the  quantized Hall conductance.

The Chern number for the two bands of the Haldane model is\cite{haldane88,lado15b}:
\begin{equation}
{\cal C}_{\pm}=\pm {\rm sign}(t_H)
\end{equation}
For the Haldane model, the dominant contribution to the Chern number in eq (\ref{Chern})  comes from the neighborhood of the $K$ and $K'$ points.  We can actually compute the Berry curvature of a massive Dirac model ${\cal H}_0+\delta {\cal H}= \hbar v_F \left(q_x \sigma_x  + \tau_z q_y \sigma_y\right) + \frac{\Delta}{2} \sigma_z$
as

\begin{equation}
{\cal  C}(\Delta,\tau_z)= \frac{1}{2}\tau_z \frac{\Delta}{|\Delta|}
\label{ChernDirac}
 \end{equation}

For the sake of the discussion we can  think of the Berry curvature as a magnetic field and the Chern number as a flux. 
From eq. (\ref{ChernDirac}) we see that in the Haldane model the sum over the two valleys leads to a ${\cal C}=1$. In contrast,
when the gap is opened by the breaking of sublattice symmetry,  as in eq. (\ref{BN}) the total Chern number vanishes.

So far, the discussion has ignored the spin degree of freedom and all our statements hold true for each spin channel separately.
Therefore, if we define a Haldane model for spinful fermions, the Chern number would be 2.   In 2004 Kane and Mele\cite{kane05} found out that the model Hamiltonian that describes graphene with intrinsic spin orbit coupling is mathematically equivalent to a {\em spin-dependent second-neighbour  Haldane coupling}:
\begin{eqnarray}
{\cal H}_{\rm KM} =
 \left(
\begin{array}{cc}
s t_{KM} g(\vec{k})  &  t\,f(\vec{k}) \\
 t\,f^*(\vec{k}) & - s t_{KM} g(\vec{k})
\end{array}
\right)
\end{eqnarray}
where $s=\pm 1$ labels the spin. 
Thus, graphene was predicted to have a  gap at the Dirac point, given by $3\sqrt{3}t_{KM}$. 
Unlike the Haldane model, the Kane-Mele model preserves time reversal symmetry and, consequently,
the total Chern number, once we sum over the spins, is zero.  Therefore,  the Kane-Mele model does not describe a Chern insulator.
Yet, the model defines a  topologically insulator,  different from vacuum,  that is characterized by a $Z_2$ topological index and the emergence of spin-chiral edge states (see next section). The magnitude 
the gap is predicted to be smaller than a fraction of meV, on account of the very small spin-orbit coupling of carbon. Therefore,
for most practical purposes, graphene can be considered gapless. Yet, the conceptual importance of this prediction was very important, as it triggered the quest of artificial
graphene realizations with this type of topological gap\cite{jotzu14}.

Once spin-degree of freedom is considered, there are at least two other mechanisms to turn graphene into a Chern insulator. The first, proposed by Z. Quiao and coworkers\cite{quiao10}, combines an off-plane Zeeman splitting, induced for instance by spin proximity with a ferromagnet with off-plane easy axis,  and a second type of spin-orbit interaction that arises in graphene when mirror symmetry is broken, the so called Rashba spin-orbit coupling.  The Zeeman splitting of the Dirac cones leads to crossing of the electron-like band of one spin channel with the hole-like band of the other, exactly at $E=0$, with coincides with the Fermi energy at half-filling.   Rashba spin-orbit coupling mixes the two spin channels, leading to a band-gap opening at the Dirac energy, that turns out to have a finite Chern number.
An alternative mechanism not relying on spin-orbit coupling
consists on using exchange interaction to a magnetic skyrmion
lattice placed underneath graphene\cite{lado15a}.

\section*{Calculation methods for  edge states in graphene-like systems}
\label{sec:2}
The calculation of edge states in graphene is not as straightforward as the calculation of bulk states on account of the reduced crystal symmetry of the problem.  The  are two  widespread methods:  1) Computation of energy bands of 1D stripes or ribbons.  2) Green's function methods.

\subsection*{Energy bands of 1D stripes}
We choose a crystal direction and a given width $W$ and define a 1D-crystal or ribbon  with a  two edges.  When the width of the ribbon is sufficiently large, compared to the penetration length of the edge states, the interactions between the edges are  negligible, allowing to study the properties of the edge states as if they were decoupled. This method is very versatile as it permits one to include all types of terms in the Hamiltonian, including a magnetic field.

\subsection*{Green's function}
In this approach\cite{lado15b}, we calculate the Green's function of a semi-infinite two dimensional crystal using the recursion method.  Using the translation invariance along the direction parallel to the edge, it is possible to write the Hamiltonian of the semi-infinite crystal as a one dimensional semi-infinite crystal whose effective Hamiltonian depends on the transverse wave vector $k_\parallel$. By so doing, the Green's function of the unit cell in the semi-infinite crystal reads:
 \begin{equation}
G (k_\parallel,E) = \left ( E - h_0(k_\parallel) - \Sigma(k_\parallel,E) + i\epsilon      \right )^{-1}
\label{green1}
\end{equation}
where $\Sigma(k_\parallel)$ is the self-energy induced by the coupling to the rest of the crystal: 
\begin{equation}
\Sigma (k_\parallel,E) = t(k_\parallel)G (k_\parallel,E) t^\dagger(k_\parallel)
\label{self1}
\end{equation}
where $h_0(k_\parallel)$ and $t(k_\parallel)$ are the intracell and intercell matrices of the Block Hamiltonian. Equations (\ref{green1}) and (\ref{self1}) define a set of non-linear coupled matrix equations that is solved by numerical iteration.   Once the surface Green's function  is derived, the density of states can be obtained through $A(k_\parallel,E)= -\frac{1}{\pi}{\rm Im}\left(G(k_\parallel,E)\right)$.  Contour plots in the $k_\parallel,E$ plane can reveal the existence of in-gap edge states.
This method avoids the inter-edge coupling issues that may arise in the method of graphene ribbons when $W$ is not sufficiently large. 

\subsection*{Analytical results}
We can derive edge states in the case of gapless graphene by looking for Bloch states that have imaginary momentum in the armchair direction, $\vec{k}=(i\kappa,  k_y)$,  and requesting they have $E=0$.   Those states would be thus evanescent in half of the plane, and itinerant along the zigzag direction.  There are two ways in which we can have a state with $E=0$: either $f(i\kappa,  k_y)=0$  and the wave function localized in sublattice $A$ or $f^*((i\kappa,  k_y))=0$ and the wave function localized in sublattice $B$.   
These two situations lead to the following equation that relates 
 $\kappa$ and $k_y$ is:
\begin{equation}
e^{\pm\frac{\sqrt{3}}{2} \kappa a}= \frac{-1}{2\cos(\frac{k_y a}{2})}
\label{eq:anal1}
\end{equation}
 Since the left hand-side of this equation is positive, for real $\kappa$, we need to have  $cos(\frac{k_y a}{2})<1$. In addition, depending on the sign of the exponent of the left hand side, this entails $|cos(\frac{k_y a}{2})|$ to be either smaller or larger than $\frac{1}{2}$. Altogether, these two conditions impose limits to the values that the transverse momentum of the edge state, $k_y$, can take:  either
 in region $\frac{2\pi}{3}<|k_y a|<\pi$, or outside of this region.    As we shall show right now, these two types of edge state correspond to two types of edge termination: zigzag, or the so called "bearded", both of them along the zigzag direction. In the zigzag (bearded) case, with edge states in (out of)  the $\frac{2\pi}{3}<|k_y a|<\pi$ region, the density of edge states per edge atom is $1/3$ ($2/3$)
 
\begin{figure}
 \centering
    \includegraphics[width=0.5\textwidth]{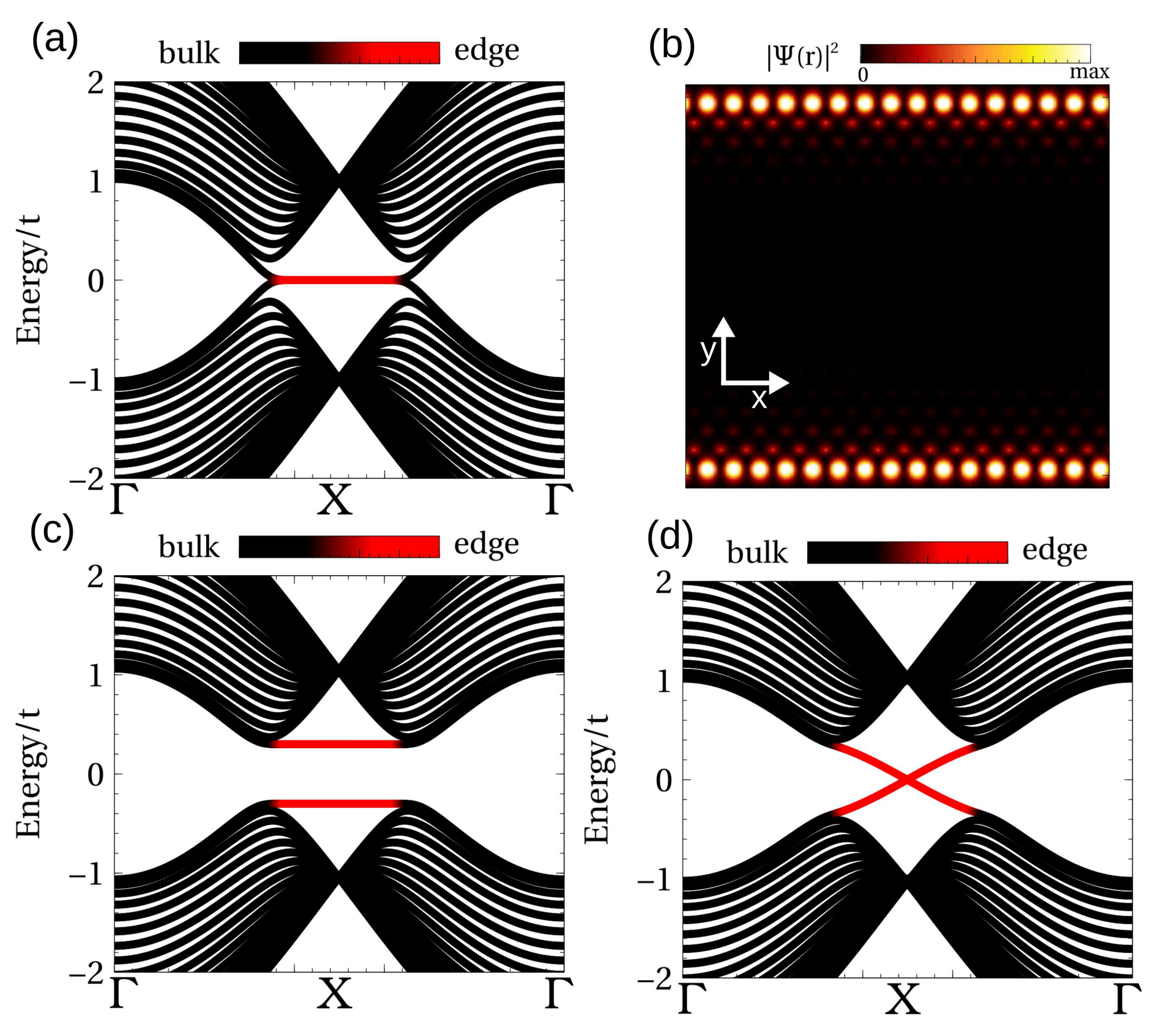}
\caption{(a)  Energy bands for zigzag ribbon in the gapless regime ($t_{KM}=t_H=\Delta=0$). The states localized at the edges are plot in red. (b)  Wave function of the one of the $E=0$ modes at the  $X$ point showing a strong localization in one of the edges.   (c) Bands for the same system adding a sub-lattice imbalance term $\Delta$ in the Hamiltonian: a band-gap opens in the edge states.  (d) Bands for the same system with a finite value of $t_H$:  the edge states,  in red color acquire now a dispersion.  The positive and negative velocity edge states are localized at opposite edges }
\label{fig:zzrib}
\end{figure}

\section*{Edge states in gapless graphene}
\label{sec:2}
 We now consider edge states in gapless graphene for zigzag edges.  The prediction of  eq. (\ref{eq:anal1}) is confirmed by inspection of the energy bands of two very wide  graphene ribbons with zigzag boundary conditions
 (see Fig. \ref{fig:zzrib}a). Most of the bands in these figures correspond to confined {\em bulk} (2D) states. Since the ribbons have two edges, there are two   flat bands with energy $E\simeq 0$.  In the case of zigzag (bearded) in the region $\frac{2\pi}{3}<|k_y a|<\pi$.  Since the extinction factor $\kappa$ goes to zero for $k_y a= \frac{\pm 2\pi}{3}$, there is a small hybrydization of the two edge states in that region, due to the finite width of the simulation ribbon. 
  For a given edge,  inspection of the wave functions show that wave functions are sub-lattice polarized at opposite sub-lattices for the two terminations, as expected from the analytical argument.  An analogous calculation with a ribbon with armchair termination does not show edge states. 
 
 Zigzag edge states can also exist in finite size graphene systems,  provided the edges are long enough.    For zigzag termination, edge states occupy one third of the Brillouin zone. Thus, there is an average of 1/3 of  edge state per edge atom.  Accordingly, graphene fragments host edge states when they have zigzag edges with at least 3 edge sites.   For a structure with $N_A$ and $N_B$ sites in the $A$ and $B$ sublattices, respectively,  the first-neighbour TB model predicts the existence of at least $|N_A-N_B|$ zero energy states.  These zero energy states, or zero modes, are sub-lattice polarized and very often these are are  localized at the edges.  Since zigzag edges have a local sublattice imbalance,   both in the zigzag and bearded terminations, 
  
 Whereas the zigzag edge states occur in the absence of a topological gap in the band structure of 2D graphene, the existence  of an edge state with momentum $k$ can still be understood in terms of   bulk-edge correspondence between the quantized value of
the Zak phase $Z(k)$, which is a Berry phase across an appropriately chosen one dimensional Brillouin zone\cite{delplace11}.

As a result of the  large density of edge states at $E=0$, which is the Fermi energy of charge-neutral graphene,  zigzag edges are prone to magnetic instabilities, on account of electron-electron interactions\cite{yazyev10}.  As a result, local moments are expected to exist at zigzag edges.  The small magnetic anisotropy and their one-dimensional character prevent the formation of long-range magnetic order and promote quantum disordered states instead.

\begin{figure}
 \centering
    \includegraphics[width=0.5\textwidth]{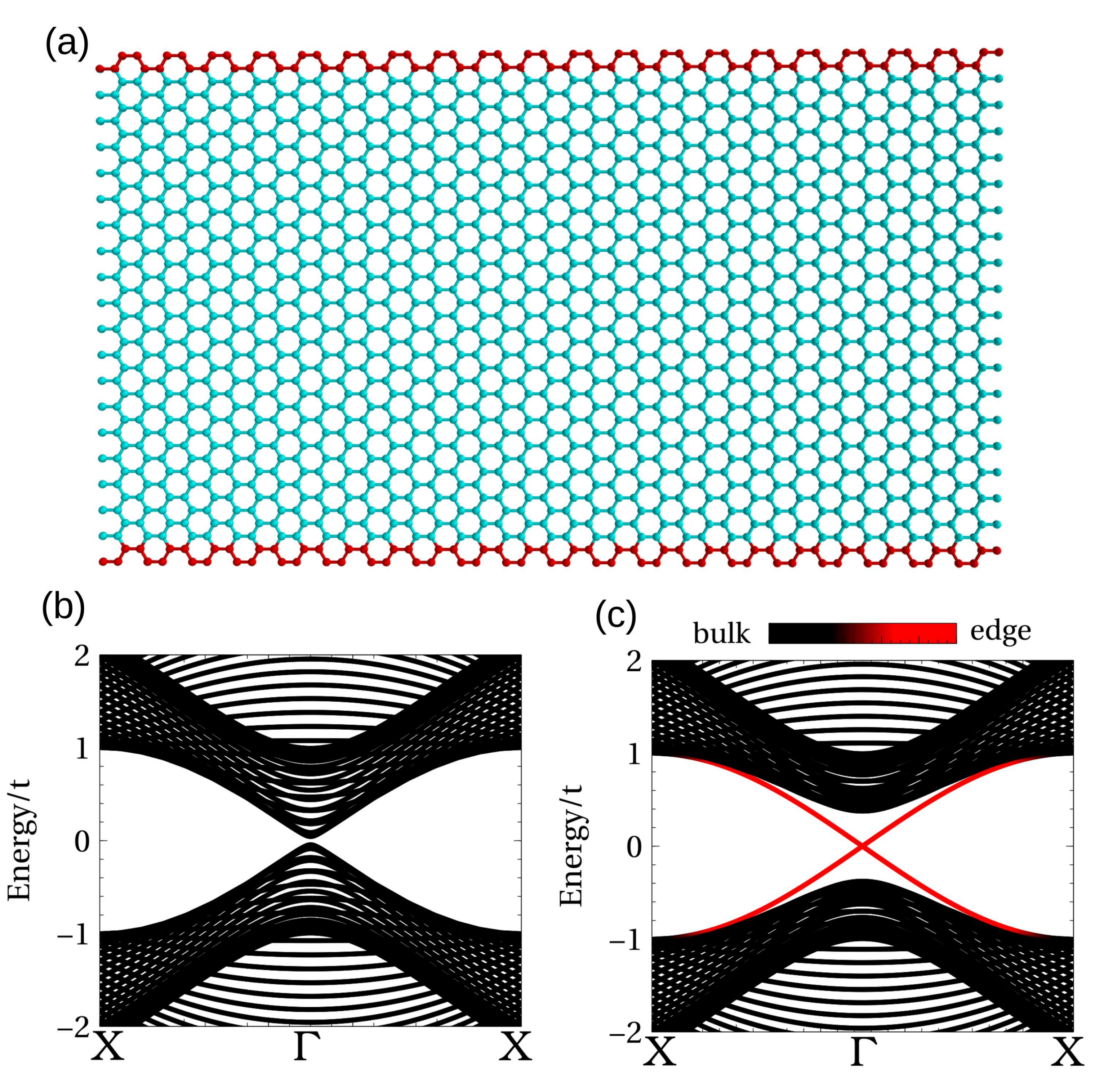}
\caption{(a)  Snapshot of the unit cell to generates the graphene ribbon, with the  
edge atoms marked in red.
(b) Energy bands for armchair ribbon in the gapless regime ($t_{KM}=t_H=\Delta=0$). The ribbon has a gaped spectrum and no edge states. (c) Bands for the same system, with  a finite value of $t_H$:  two dispersive edge states, in red color appear in the gap. This supports the statement that topological edge states appears in all boundaries, regardless of their geometry }
\label{fig:acrib}
\end{figure}

 \section*{Edge states in topological graphene}
 We now consider edge states  associated to the various topological phases of graphene.  There are two  main differences with the edge states of gapless graphene. First, edge states occur for all edges. Second, they are dispersive.   Third, for a given edge,  the group velocity of edge states takes one sign only: they are {\em one-way states}. 
 

 We discuss three different topological phases in graphene: the Quantum Hall phase, the Chern insulator phases,  and the Quantum Spin Hall phase.
 
 \subsection*{Quantum Hall phase}
Application of a magnetic field perpendicular to a two dimensional electron gas (2DEG) results in the Landau quantization of the energy spectrum. Classically,  electrons follow closed orbits. Quantum mechanically,   electrons occupy finite size states, with a discrete energy spectrum $\epsilon_n=\hbar\omega_0(n+\frac{1}{2})$, where $\omega_c=\frac{eB}{m} $  is the cyclotron frequency, $n$ are integer number greater than 0. These are  so called Landau levels.  Landau levels have a macroscopic degeneracy, associated to the location of the classical orbits across the plane. The density of such orbits, and thereby the degeneracy of the Landau levels, are determined by the magnetic field. When the density of electrons matches the density of states per Landau Level, the occupation of Landau levels at zero temperature is either complete or null: this is a Quantum Hall insulator, with the Fermi energy lying in the gap between Landau levels. 
Unlike normal insulator, Quantum Hall insulators have in-gap edge states and their  Hall conductance is finite and quantized\cite{klitzing80} in units of $e^2/h$. Edge states are often interpreted in terms of semiclassical skipping orbits: scattering of electrons with the edge prevents them from completing closed loops\cite{halperin82}.  However,  the robustness of edge states and their contribution to Quantum Hall conductance quantization lies ultimately in the topological nature of bulk states\cite{thouless82}, and not on the specifics of the edges. 

The Quantum Hall phase  picture just described holds for graphene with several differences. For Dirac electrons, instead of just one,  there are two sets of Landau levels  with electron-hole symmetry, one for electrons, one for holes\cite{neto09}. Thus, for the valley $\tau=\pm 1$, there are two sets of Landau Levels  $E_n=\pm \frac{\hbar \omega_{0g}}{2}\sqrt{ n+\frac{1-\tau}{2}}$ where $\omega_0=\frac{2 v_F}{\ell_B}$, $\ell_B=\sqrt{\frac{\hbar}{eB}}$ and $v_F$ is the velocity of the Dirac electrons.  In addition to these four sets of finite energy Landau levels (two per valley),
there are two $E=0$ Landau levels, one per valley.  

These Landau levels, and the edge states, can be described with
the same tight-binding model that describes a finite width graphene ribbon, including the effect of the magnetic field on the orbital motion of the electrons by using the Peierls substitution, where the hopping between sites $\alpha$ and $\beta$ is replaced by:
\begin{equation}
t_{\alpha\beta}\rightarrow t_{\alpha\beta}e^{i \Phi_{\alpha,\beta}}
\end{equation}
where $\Phi_{\alpha,\beta}=\int_{\vec r_\alpha}^{\vec r_\beta} \vec{A}\cdot d\vec{r}$ is the Peierls phase of  the vector potential $\vec{A}$   associated to the magnetic field $\vec{B}=\vec{\nabla}\times\vec{A}$. For a constant magnetic field we can choose the vector potential as $\vec{A}=B(-y,0,0)$, so that 
translational invariance is broken along  the $y$   direction, taken to be perpendicular to the boundaries of the ribbon. Therefore, the energy states can be classified in one-dimensional energy bands. 

\begin{figure}
 \centering
    \includegraphics[width=0.5\textwidth]{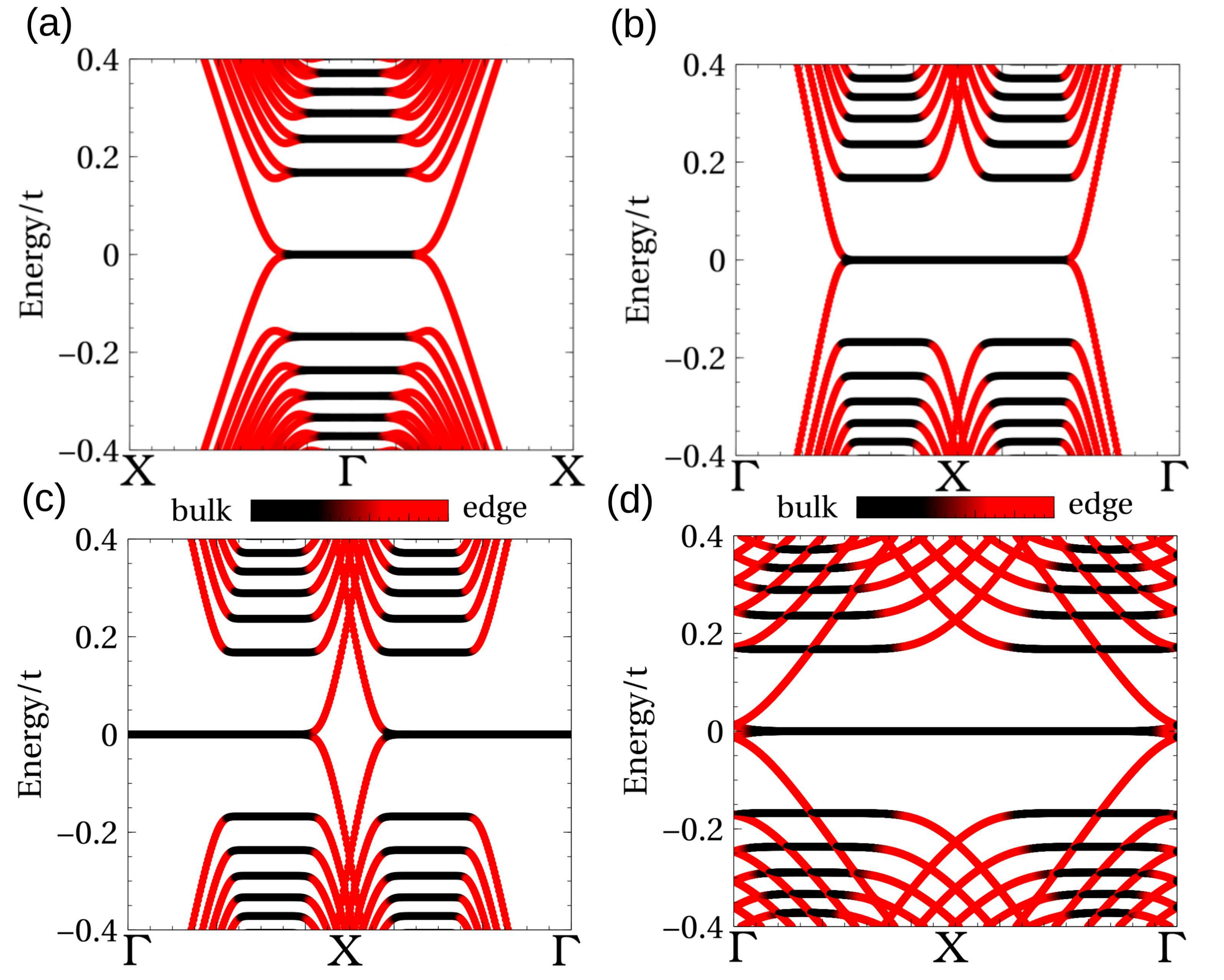}
\caption{Energy levels for armchair (a)
zigzag (b) bearded (c) and chiral (d) graphene ribbons under the influence of a magnetic field $B$ applied perpendicular to the ribbon. For the sake of clarity, spin effects are not included in the calculation.
Two types of states are shown. Dispersive edge states  in red and dispersionless states in black, corresponding to Landau levels localized away from the edges.  In addition, for the zigzag case, there are $E=0$ flat edge states, that were already present for $B=0$. 
In the zigzag, bearded and chiral geometry it is apparent the presence of two sets of Landau levels, corresponding to the two valleys. In the armchair geometry the valley  degeneracy manifested as a twofold degeneracy of the Landau Levels. 
\label{fig:LL}
}.
\end{figure}

In figure  \ref{fig:LL} we show the bands,  for  graphene ribbons both with armchair, zigzag, bearded and chiral termination.  All bands have two regions, a dispersionless region that corresponds to Landau Levels, localized away from the edges, and a dispersive region associated to edge states.  The calculation shows that finite energy levels have a twofold degeneracy, broken a the edges, associated to the valley degree of freedom. In contrast, the $E=0$ LL is formed of a duet  of non-degenerate electron-type and a hole-type Landau level. 
In the figure we also show the  how  that dispersive states correspond to edge states and their velocity is opposite for opposite edges. 

Both for the 2DEG and for graphene,  the Quantum Hall conductance is given by the TKNN formula\cite{thouless82} so that it is apparent the ultimate origin of its quantization is the topological nature of the band gap.

\begin{figure}
 \centering
    \includegraphics[width=0.5\textwidth]{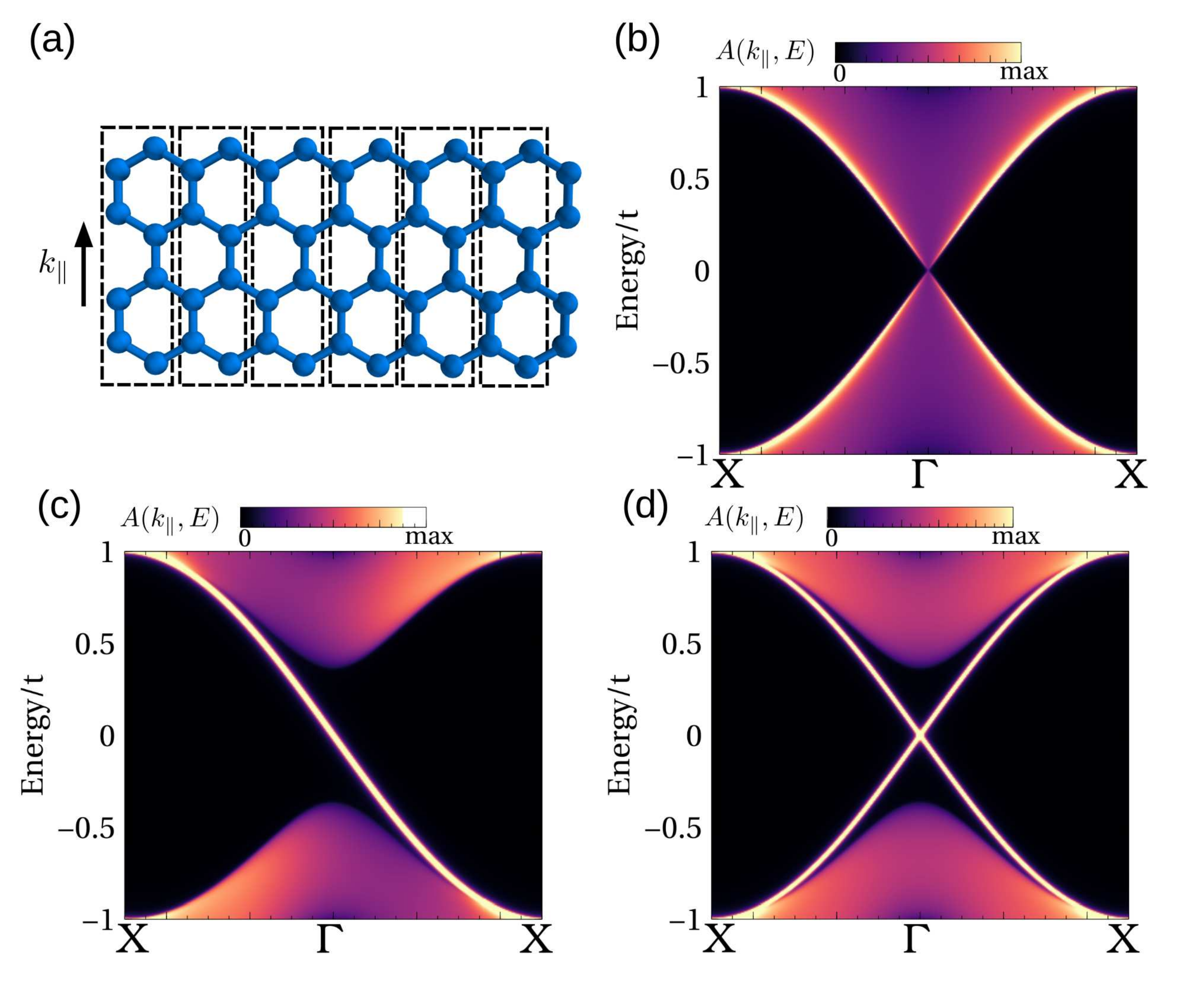}
\caption{Schematic of the Green's function method for an armchair edge (a), and
edge spectral function for a gapless armchair edge (b),
with Haldane coupling (c), and with Kane-Mele coupling (d).
}
\label{fig:green}
\end{figure}

 \subsection*{Edge states in the Chern insulator  and QSH phases}
 We now discuss the properties of the edge states of the Chern insulating phase, as described by the 
 Haldane model, and the Quantum Spin Hall (QSH) phase, as described by the Kane and Mele model.  In both cases, there are dispersive in-gap edge states, regardless of their crystal orientation (see Figs. \ref{fig:zzrib}d, \ref{fig:acrib} and \ref{fig:green}(c) for the Haldane model, and Fig. \ref{fig:green}(d), all of them for armchair edges).  In the case of figure  \ref{fig:green} we show the local-density of states,  computed with the Green's function method, resolved in transverse-momentum, for at the armchair boundary of a semi-infinite honeycomb lattice for three models: gapless graphene (panel \ref{fig:green}b), Haldane model (panel \ref{fig:green}c), and Kane-Mele model (panel \ref{fig:green}d).     At a given edge, the sign of the slope of the in-gap edge state in the Haldane model is given by the sign of $t_H$. Since  the Kane-Mele model is equivalent to two copies of the Haldane model where the sign of the coupling is given by the spin projection along the off-plane direction, it is immediately apparent that each edge host two edge states, one per spin channel, with opposite velocities.  This is clearly seen in figure \ref{fig:green}(d), obtained for a single edge using the Green's function method. 
 
\begin{figure}
 \centering
    \includegraphics[width=0.5\textwidth]{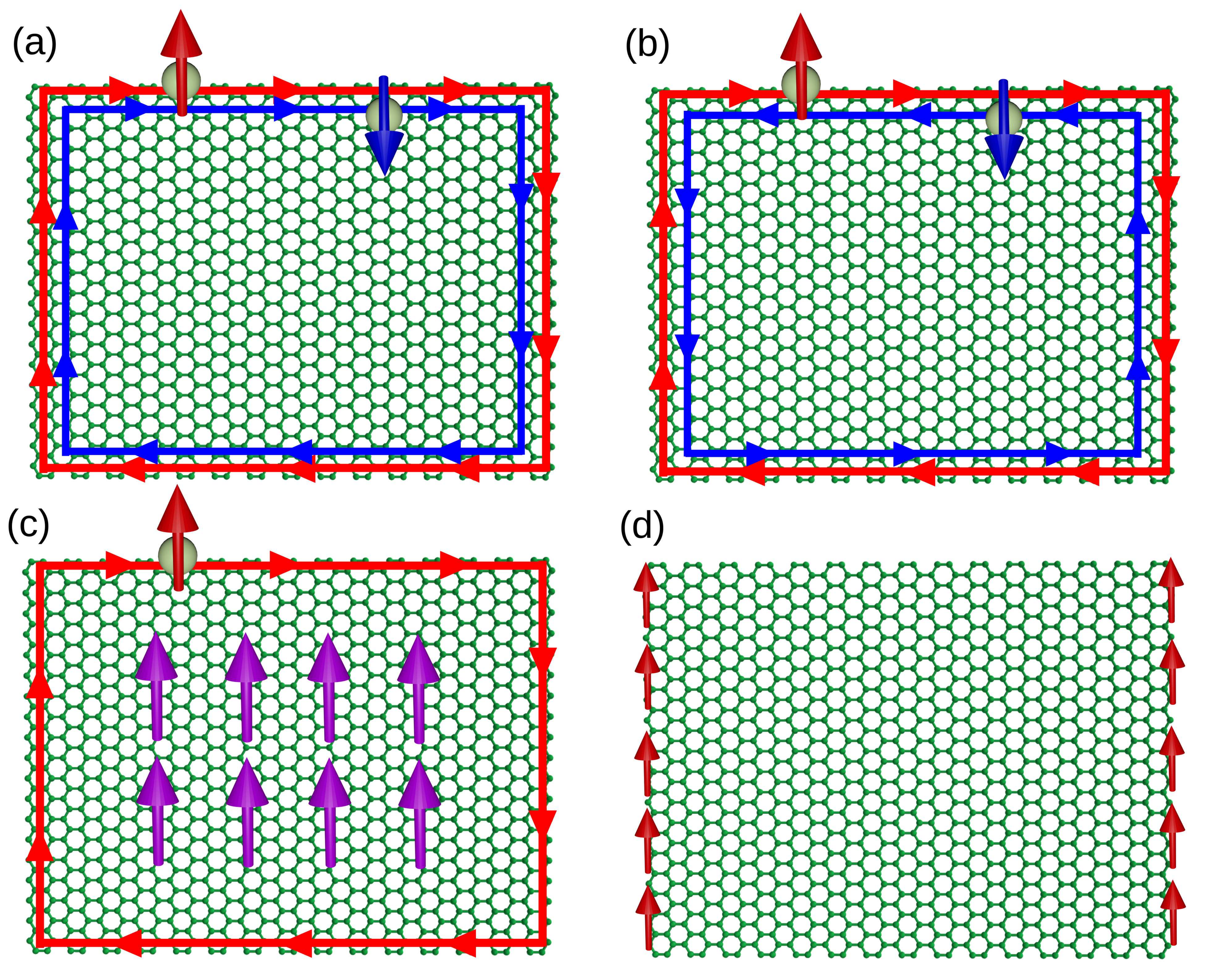}
\caption{Schematic of quantum Hall edge states (a), quantum spin Hall edge states (b), quantum anomalous Hall edge states (c) and zigzag edge states (d).
}
\label{fig:summary}
\end{figure}

 A graphical summary of the different types of edge states is shown in figure \ref{fig:summary}.  For the topologically equivalent Quantum Hall and Chern insulating phases,  the edge states circulate around the sample with sense of rotation dictated by the term in the Hamiltonian that breaks time reversal symmetry, either the external magnetic field in the Quantum Hall case, or the Haldane term in the Haldane model.  Edge states for these two phases can not undergo intra-edge elastic backscattering as there are no states  with opposite momentum and the same energy available. This can be connected both with the perfect quantization of their Hall conductance and their longitudinal conductance.  In the case of the Kane-Mele model the edge states feature counter-propagating states with opposite spin projections. Therefore, elastic backscattering is permitted in the presence of spin-flip perturbations and not permitted if time-reversal symmetry is preserved.

\section*{Importance of Edge states}

 Edge states have attracted interest for several reasons. In the case of non-dispersive edge states,  it is expected that Coulomb repulsion will promote the emergence of local moments, on account of the  that flat bands at the Fermi energy. There is a large body of research papers, mostly theoretical, devoted to the  study of magnetism in zigzag edge states in graphene.   Intra-edge interactions are expected to be ferromagnetic, at least when interactions are short-ranged (Hubbard type). However, long-range ferromagnetic order in spin-isotropic systems is not stable in 1D, on account of quantum and thermal fluctuations. Therefore, zigzag edges are expected to host fluctuating local moments, and behave as paramagnets with an enhanced spin suceptibility.
 
 Dispersive edge states feature by the topological phases attract also enormous interest. The notion of {\em one way} states is of course very appealing, not only for electrons, but for other excitations. For instance, the prediction of topological magnons in 2D ferromagnetic materials with honeycomb lattices implies the existence of in-gap {\em one way} magnon states that may find some application in magnon-based spintronics. The perfect quantization of Hall conductance of the Quantum Hall phase\cite{} is used to define the standard of the Von Klitzing constant\cite{klitzing80}, $h/e^2$. A holy grail in quantum materials is the  quest of a Chern insulator that features quantized  Hall conductance without an external magnetic field. 

\section*{Summary}
We have presented the concept of graphene-like systems and we have discussed their edge states. We have stressed the interplay between the existence, or else, of a topological gap in the energy bands of the two-dimensional honeycomb system and the properties of the edge states. Specifically, for the non-topological gapless graphene phase, edge states exist only along some crystallographic directions, most notably the zigzag direction and they form flat bands, prone to magnetic instabilities. In contrast,  for topological phases, edge states exist in all edges, regardless of crystallographic orientation, and they have a finite velocity and topological protection against back-scattering.

{\em Acknowledgements }
J.F.R.  acknowledges financial support from 
 FCT (Grant No. PTDC/FIS-MAC/2045/2021),
 SNF Sinergia (Grant Pimag),
FEDER /Junta de Andaluc\'ia --- Consejer\'ia de Transformaci\'on Econ\'omica, Industria, Conocimiento y Universidades,
(Grant No. P18-FR-4834), 
Generalitat Valenciana funding Prometeo2021/017.
and MFA/2022/045 and funding from MICIIN-Spain (Grant No. PID2019-109539GB-C41).
J.L.L. acknowledges 
the computational resources provided by
the Aalto Science-IT project, and
the 
financial support from the
Academy of Finland Projects No. 331342 and No. 336243
and the Jane and Aatos Erkko Foundation.

\bibliography{biblio}
\end{document}